\documentclass{aa}  
\usepackage{hyperref}
\hypersetup{colorlinks, allcolors=blue}
\usepackage{graphicx}    
\usepackage{color}  
\usepackage{txfonts}
\usepackage{inputenc}
\usepackage{multicol}
\usepackage{multirow}
\usepackage{longtable}
\usepackage{subcaption}  
\usepackage{siunitx}
\usepackage{amsmath}
\usepackage{xcolor}
\usepackage{booktabs}
\usepackage{color} 

\usepackage{algorithm} 
\usepackage{algpseudocode} 
\begin{document}

   \title{Towards a complete picture of the Sco-Cen outflow}


   \author{M. Piecka\inst{1}
          \and
          S. Hutschenreuter\inst{1}
          \and
          J. Alves\inst{1,2}
          }

   \institute{University of Vienna, Department of Astrophysics,
              T\"urkenschanzstrasse 17, 1180 Vienna, Austria\\
              \email{martin.piecka@univie.ac.at}
         \and
             University of Vienna, Research Network Data Science at Uni Vienna, Kolingasse 14-16, 1090 Vienna, Austria}

   \date{Received X XX, XXXX; accepted X XX, XXXX}


  \abstract{

Previous studies have shown strong evidence that the Sun is crossing an outflow originating from the Sco-Cen OB association. Understanding this outflow's origin and structure illuminates how massive star formation shapes the interstellar medium (ISM) and helps predict future Galactic conditions affecting our Solar System.
We analysed H~\textsc{i} emission and optical ISM absorption lines towards 47 early-type stars around the Upper~Sco region to refine the map of the Sco-Cen outflow.
Combined with data for nearby stars, we find that the outflow has at least two components: a faster, low-density component traced by Ca~\textsc{ii}, and a slower, possibly lower-density component traced by Mg~\textsc{ii} and Fe~\textsc{ii} in the UV that is passing through the Earth. A constant flow model successfully describes both components with $(l,b,|\vec{v}|) = (335.4^{\circ}, -6.8^{\circ}, 14.0 \,\,\textrm{km}\,\textrm{s}^{-1})$ and $(305.5^{\circ}, +17.6^{\circ}, 21.2\,\,\textrm{km}\,\textrm{s}^{-1})$, respectively. The origin of the faster component points towards the Sco-Cen 15 Myr population, which is consistent with the origin of the slower component within 2~$\sigma$. A simple model comparison indicates that a constant flow is favoured over a spherical flow geometry, implying an extended distribution of feedback sources within Sco-Cen. We also found that a poorly studied 25 pc long H~\textsc{i} cloud at a distance of 107 pc belongs to the established Sco-Cen flow.  
  
  }


   \maketitle
%
\section{Introduction}\label{section:1}

The Scorpius-Centaurus OB association (Sco-Cen) is known to produce blue-shifted interstellar absorption lines in the spectra of nearby stars, suggesting an outflow of material from its young stellar population \citep[e.g.,][]{Hobbs1969-os,Frisch1986-lv}. Since then, research in the local interstellar medium (ISM) has revealed several warm, low-density clouds within 30 pc, whose kinematics suggest an association with this outflow \citep[e.g.,][]{1995SSRv...72..499F,2002ApJ...574..834F, 2008ApJ...673..283R,2011ARA&A..49..237F}. Likely related, the detection of $^{60}$Fe in deep-sea sediments suggests a recent influx of supernova ejecta into the Solar System, likely originating from Sco-Cen or the Tucana-Horologium association \citep{2002PhRvL..88h1101B, 2005ApJ...621..902F, 2016Natur.532...69W, 2018AN....339...78H, 2023A&A...680A..39S}.

As noted in \citet{1991A&A...247..183C}, there is evidence for a gaseous outflow from Sco-Cen that encompasses the Ophiuchus and the Lupus clouds.  However, due to a limited number of sightlines probing the ISM, they were unable to identify any specific spatial sub-structure within the flow, though their high-resolution spectra did reveal kinematic sub-structure. Such spatial and kinematic sub-structures are expected due to the complex nature of outflows, as demonstrated in simulations \citep[e.g.,][]{2013MNRAS.431.1337R}. More recently, \citet{2008A&A...483..471N} conducted a detailed study of the spectrum towards HD 102065, identifying kinematic sub-structure with the most negative component at -20~km\,s$^{-1}$ (w.r.t. the local standard of rest, LSR), similar to the most negative radial velocities found by \citet{1991A&A...247..183C}.

Sco-Cen is the closest OB association to Earth, containing stars as old as $\sim$20 Myr and as young as protostars \citep{2012ApJ...746..154P,2019A&A...623A.112D,2021ApJ...917...23K,2022A&A...667A.163M,2023A&A...677A..59R}. Recently, \cite{2023A&A...678A..71R} constructed a high-resolution star formation history map of Sco-Cen showing the existence of a dominant formation peak about 15 Myr ago, where most of that stars and clusters were formed.  This work also revealed chains of ordered star-forming regions, likely formed from the feedback of Sco-Cen's massive stars over the last 10 Myr \citep{2023A&A...679L..10P}. This feedback flow is likely to also be responsible for the enrichment of star-forming clouds in Ophiuchus with short-lived radionuclides like $^{26}$Al  \citep{Forbes2021-dn}.

\begin{figure*}
 \centering
 \includegraphics[width=2\columnwidth]{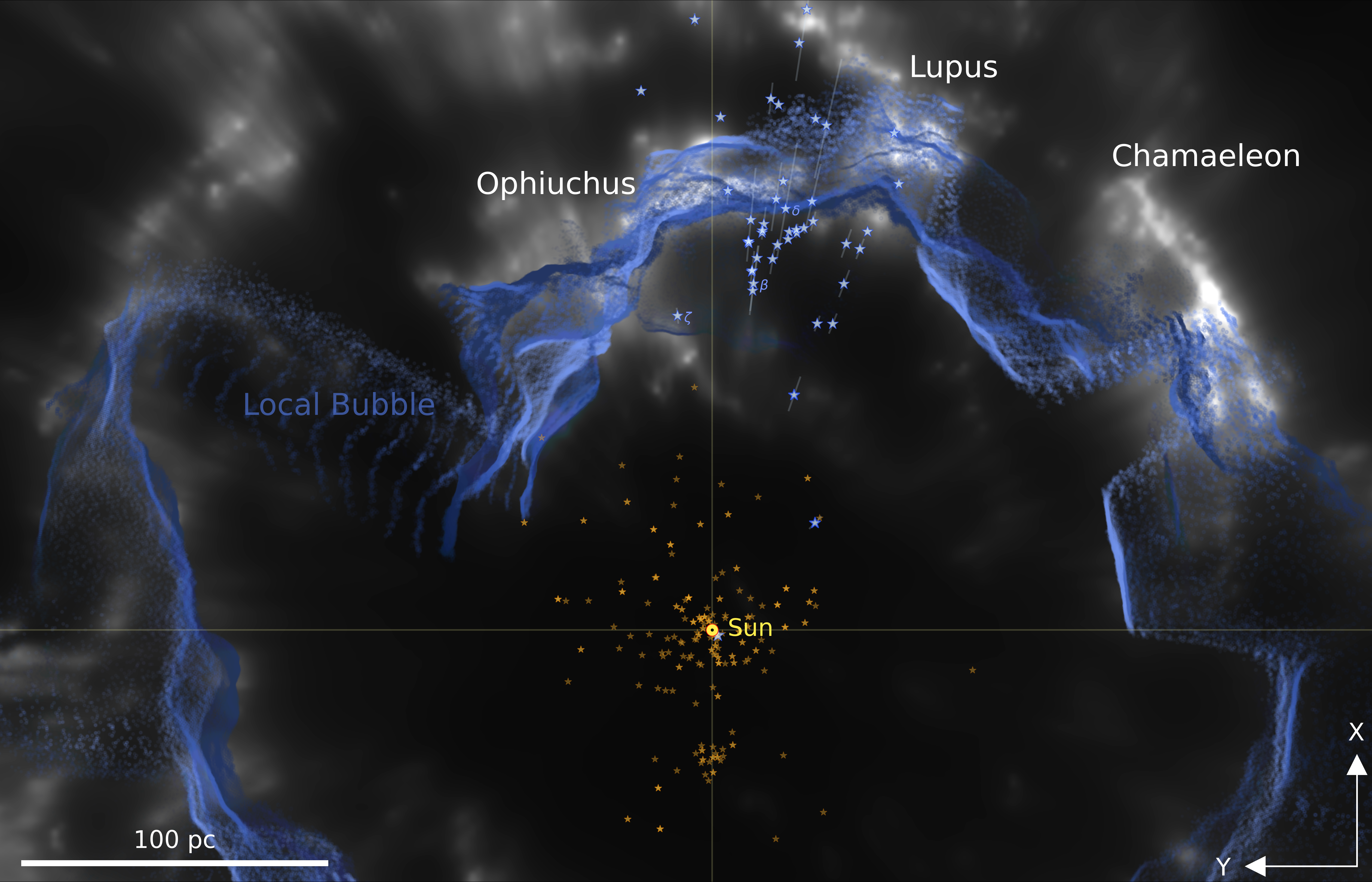}
 \caption{Map of the studied region between the Sun and Sco-Cen. The grey-shaded areas represent the dust map from \citet{GordianDustMap} projected onto the $XY$-plane, while the blue-shaded areas (brighter closer to the Galactic plane) display the Local Bubble map \citep[][$-50 < Z < 50$ pc]{TheoLocalBubble} based on the same dust map. The smaller orange stars show the positions of the objects from \citet{2008ApJ...673..283R}, where brighter orange stars represent measurements with multiple velocity components. The large blue stars indicate the positions $\left(d = \varpi ^{-1}\right)$ of our calcium-probing (mostly B-type) stars, with lines indicating distance uncertainties. The two arrows in the bottom right corner show the directions towards the Galactic centre (up) and the Galactic rotation (left). The bar in the bottom left corner (length of 100~pc) serves as a measure of the distance scale in the map.}
 \label{fig:stage}
\end{figure*}

In this Paper, we attempt to connect previous observations of the Sco-Cen outflow with archival data to begin an investigation on the structure of the Sco-Cen flow. The region of interest is the thin ISM between Sco-Cen and Earth, as represented in Fig.~\ref{fig:stage}. We make use of the combination of spectral information obtained with ESO telescopes to map ISM absorption lines, combined with astrometric Gaia Data Release~3 \citep{2021A&A...649A...1G,2023A&A...674A...1G} and Hipparcos \citep{1997A&A...323L..49P,2007A&A...474..653V} to constrain ISM distances. We also use H~\textsc{i} data \citep{2016A&A...594A.116H} in our analysis. Unless stated otherwise, all velocities are presented in LSR.

This study aims to illuminate the interactions between the massive stars and the gas surrounding the Sco-Cen association and improve our understanding of the source of the feedback that shapes the local ISM, contributing to a better understanding of the environmental conditions affecting the Solar System.

\section{Archival data}\label{section:2}

We make use of the spectra available in the ESO archives. Two instruments are associated with data products that have wavelengths already shifted to the barycentric rest frame: FEROS \citep[$\textrm{R}=48\,000$,][]{1999Msngr..95....8K} and HARPS \citep[$\textrm{R}=80\,000/115\,000$,][]{2003Msngr.114...20M}.
Since FEROS observations cover a larger number of Sco-Cen stars, specifically within and around the Upper~Sco population, we use only these spectra for extracting information about spectral lines.
We use HARPS only for a visual inspection of spectral lines at a higher resolution. UVES \citep{2003Msngr.114...10B} is used for investigating an interstellar titanium line in the near-UV.

Our primary aim is to make use of two interstellar spectral lines within the $3800-8000$~{\AA} region: Ca~\textsc{ii} K~line (3933.66~\AA) and K~\textsc{i} line (7698.96~\AA). Ca~\textsc{ii} and K~\textsc{i} are both doublets, but we are forced to ignore Ca~\textsc{ii} H~line at 3968~{\AA} and K~\textsc{i} at 7665~{\AA} due to blending with a stellar (Balmer-$\epsilon$) and a telluric feature (O$_2$ A-band), respectively. The Ca~\textsc{ii} H~line is used to validate the source of features around 3933~{\AA} attributed to calcium, together with the Ti~\textsc{ii} line at 3384~{\AA}, which is considered as a tracer of comparable ISM conditions as those probed by Ca~\textsc{ii} \citep[for example][]{1997ApJS..112..507W}.

To be able to detect the interstellar calcium line, the spectrum of the studied star must contain a very weak (or ideally no) stellar feature at this wavelength. This forced us to focus on studying only lines of sight towards hot-type stars (O, B, possibly early A), with the spectral class cut-off depending on the projected rotational velocity of the star. We identify 45 feasible targets with available FEROS spectra (see Appendix~\ref{section:A}). These objects are located at distances between 100 and 300~pc. In the sky, the targets are spread around the $\rho$~Oph region within the radius of about $17^{\circ}$. The central regions of this area are more densely covered when compared to the outskirts, which is the result of an observational bias -- the hot stars in Upper~Sco are usually located closer to the high-extinction star-forming region. To investigate the detailed sub-structure of calcium \citep[for example][]{1996ApJS..106..533W}, we also extracted the HARPS spectra for 9 stars of our sample. Additionally, the spectra of HD~158427 ($\alpha$~Ara) and HD~116658 (Spica) were also obtained to check the profiles of optical interstellar lines towards B-type stars at lower distances.

Identifying an outflow within 70~pc from the~Sun using hot-type stars is impossible due to the lack of such objects. Instead, one needs to rely on different methods. \citet{2008ApJ...673..283R} focused on later-type stars (mostly cooler than B) at closer distances, bridging the gap between more distant and nearby probes of the ISM. In this work, we use their published heliocentric radial velocities extracted from \textit{HST} UV spectra.

To investigate a potential link between the dense ISM around the studied OB stars and the flow traced by calcium, we incorporate additional data into our analysis. We use the HI4PI survey \citep{2016A&A...594A.116H}, which offers the most comprehensive all-sky coverage of neutral hydrogen (H~\textsc{i}), to determine the velocities of denser structures. This survey offers a spatial resolution of $\ang{;16.2;}$ and a kinematic resolution better than 2 km,s$^{-1}$. Finally, we use the 3D dust map of \cite{GordianDustMap}, which will play a critical role in allowing a distance determination towards structures identified in H~\textsc{i}.

We note the availability of the Ca~\textsc{ii} (and the Na~\textsc{i}) map from \citet{2010A&A...510A..54W}. However, we find that this map is of a limited use, as it does not distinguish between the velocity components seen in the profiles of the spectral lines.

\section{Analysis of Ca~\textsc{ii} and K~\textsc{i} spectral lines}\label{section:3}

The continuum normalisation was performed by masking the interstellar and stellar features and fitting a cubic spline, similar to the approach described in \citet{2020A&A...641A..35S}. Afterwards, stellar/telluric features were fitted with a combination of several generalised Gaussians and subtracted from the spectrum, while keeping the interstellar line of interest masked. The reduced spectrum was finally fitted with a combination of standard Gaussians. All of the components of the observed spectrum (continuum, stellar/telluric features, interstellar lines) were fitted using \texttt{scipy.optimize.curve\_fit}.

At least two Gaussian components are required to properly fit the profiles of the Doppler-splitted calcium line in the spectra of our targets. We chose not to fit more complicated profiles due to the limited resolution of FEROS. Detection of the same splitting in the profiles of the Ca~\textsc{ii} H~line and the near-UV Ti~\textsc{ii} line (if a UVES spectrum is available), we can confirm that the complex profile of the calcium line is the result of Ca~\textsc{ii} Doppler-splitting and not an overlap of unrelated features. The effect of splitting was observed in potassium only in the case of HD~142184, where a blue-shifted component ($\Delta \textrm{RV} \approx -7$~km\,s$^{-1}$) appears in both of the K~\textsc{i} doublet lines. For this specific case, the blue-shifted component was masked during the fitting procedure. The resulting radial velocities are presented in Table~\ref{table:A} -- for an easier comparison with the literature, RVs were shifted to the LSR using \texttt{astropy}, where $\left( U_{\odot},V_{\odot},W_{\odot} \right) = \left( 11.10, 12.24, 7.25 \right)$~km\,s$^{-1}$ was obtained by \citet{2010MNRAS.403.1829S}. Additional information about the extracted values and the corresponding errors is provided in Appendix~\ref{section:A}.

We find no prominent intervening cloud in front of Spica and $\alpha$~Ara. This is revealed by the lack of potassium absorption towards these two lines of sight. The spectra of both stars show two calcium components -- one located at $\sim -7$~km\,s$^{-1}$, and an offset component that is either red-shifted (Spica) or blue-shifted ($\alpha$~Ara). 

The distribution of RVs (Fig.~\ref{fig:A}) of the blue and the main component of calcium are centred at around $-15$ km\,s$^{-1}$ and $+3$ km\,s$^{-1}$, respectively. Comparing with the CO velocity of the $\rho$~Ophiuchus cloud \citep[between $+1$ and $+4$ km\,s$^{-1}$, see][]{2001ApJ...547..792D}, we find that the velocity distribution of the main calcium component fits well with the velocity of the Ophiuchus complex. We also note a relatively strong one-to-one relation between the RVs of potassium and calcium main (Pearson coefficient $\rho=0.78$). The identified velocity of the blue component coincides with the velocity of the outflow, as suggested in the literature (see Section~\ref{section:1}). HARPS spectra reveal that the blue velocity component in not a single line but consists of about two or three sub-components (see Fig.~\ref{fig:B} for an example) that are unresolved in most of the FEROS spectra (see Appendix~\ref{section:B}).

\section{Modelling the Sco-Cen outflow}\label{section:4}

\begin{figure*}
 \centering
 \includegraphics[width=2\columnwidth]{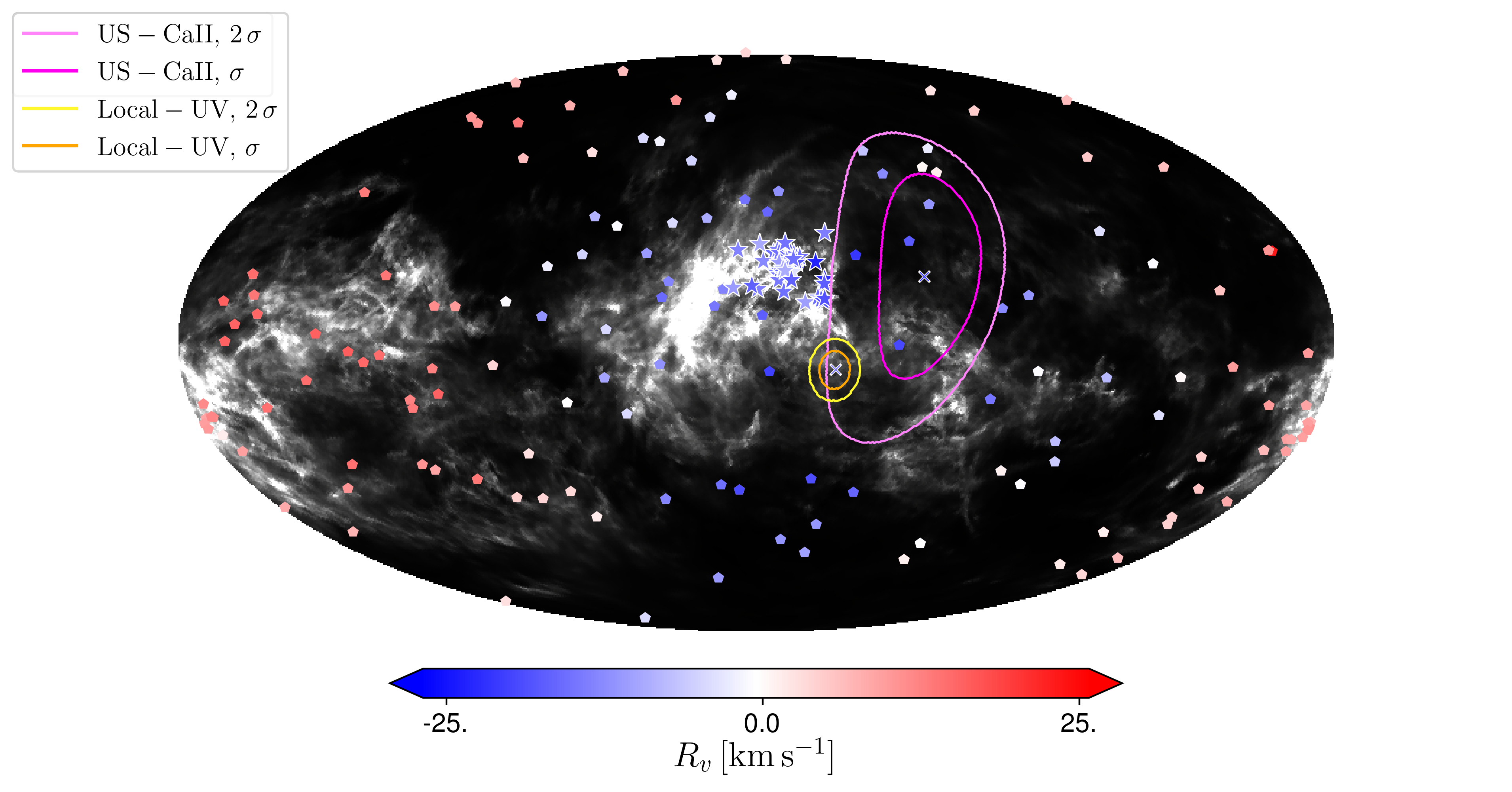}
 \caption{Marginal posterior distributions of the outflow direction in the sky. The image is centred on $(l,b) = (0^{\circ},0^{\circ})$. We show the contours of the 1 and 2~$\sigma$ quantiles: orange-yellow colours for Ca~\textsc{ii}, magenta colours for the local flow data from \citet{2008ApJ...673..283R}. We highlight the respective posterior means with large blue cross symbols. 
 The positions of the stars are shown as pentagons (local flow) and stars (Ca~\textsc{ii}), with the amplitude of the radial velocity of both data and posterior mean indicated by the blue-red colour-map.
 The background gray-scale shows the \citet{GordianDustMap} dust map, integrated up to 400 pc.}
 \label{fig:sky}
\end{figure*}

By taking into account the results from \citet{2008ApJ...673..283R} and the results of our analysis of the Ca~\textsc{ii} lines, we have enough information to make a statement about the structure of the observed flow in the ISM. We attempt to estimate the 3D velocity vector field using simple flow models. In what follows, we treat the UV spectral lines probing the local ISM separately from the Ca~\textsc{ii} lines observed towards the Upper~Sco.

The first model used to determine the origin of the flow and its amplitude is a constant (and uniform) vector field, or a constant flow for brevity. This model is parameterised by the vector $\vec{v} = \left(v_x, v_y, v_z\right)$, defined in the same coordinate system as the map in Fig.~\ref{fig:stage} ($z$ points out of the plane, Sun is in the origin).
From this, the radial velocity of the flow between us and a star can be calculated by projecting this vector on the respective normalised position vector in heliocentric coordinates.
This model is likely the simplest model to fit an outflow, and can at best constrain the large scale velocity amplitude and angular origin of the flow.

The second model is an infinitely thin expanding spherical shell, parameterised by the position of its origin in spherical coordinates at a distance $r$, $\vec{o} = \left(l, b, r\right)$, the sphere radius $R$, and the velocity amplitude perpendicular to the surface of the shell, $v_\perp$.
Such a model should work well in the context of past supernova events occurring in Sco-Cen \citep[for example][]{2001ApJ...560L..83M,Breitschwerdt2016-le}. 
The respective radial velocity in the line of sight towards each star is calculated by projecting the normalised position vector onto the normal vector on the sphere at the intersection of the line of sight of the star with the sphere.
While this model is slightly more complex than the simple constant flow, it has the advantage of having the ability to constrain the distance to the origin of the flow.
We only use this model for the Ca~\textsc{ii} data, as the data set from \citet{2008ApJ...673..283R} is difficult to fit with a thin expanding shell. This is due to the fact that distinguishing between the two models becomes impossible on sufficiently small scales (local flatness) given the magnitude of the measurement uncertainties. Assuming the diameter of the local complex of clouds to be 30~pc \citep{2008ApJ...673..283R}, a distance of 145~pc from the source \citep{2018A&A...619A.120K}, and an expansion velocity of about 25~km\,s$^{-1}$ \citep[][and citations therein]{2008ApJ...673..283R}, we estimate a deviation of maximum 0.5~km\,s$^{-1}$ between the two models -- a value below the limit set by the turbulence \citep{2022AJ....164..106L} and the expected RV errors \citep{2008ApJ...673..283R,2022AJ....164..106L}.
Furthermore, previous works already showed that a simple cosine model fits well the relation between the measured RVs and the angular separation of the line of sight from the coordinates of the flow's origin \citep[for example][]{1991A&A...247..183C,2011ARA&A..49..237F}.

To fit the data from \citet{2008ApJ...673..283R}, we make use of all of their presented radial velocities. Based on their description, we assume a constant error of 3~$\mathrm{km\,s^{-1}}$ for each data point. Unlike in \citet{2008ApJ...673..283R}, we do not aim to analyse the sub-structure of the local ISM but rather focus on the mean kinematic properties. The nested sampling algorithm \citep{2004AIPC..735..395S, 10.1214/06-BA127} is used to constrain the posterior distributions of the respective model parameters and choose widely uninformative prior distributions for all parameters, summarised in Table~\ref{table:C1}. 

\begin{table}
\caption{\label{table:C1} Priors for the two models used to fit the radial velocity data.}

\centering
\begin{tabular}{l l r }
parameter &  prior dist. & hyperparameters  \\
\hline\hline
\textbf{Con. flow model}  \\
$v_x$ [km\,s$^{-1}$] & normal & $(\mu, \sigma) = (0, 30)$ \\
$v_y$ [km\,s$^{-1}$] & normal & $(\mu, \sigma) = (0, 30)$ \\
$v_z$ [km\,s$^{-1}$] & normal & $(\mu, \sigma) = (0, 30)$ \\
  \hline\hline  
\textbf{Thin shell model}  \\
$d_l$ & uniform & $[-\ang{180}, +\ang{180}]$ \\   
$d_b$ & uniform & $[-\ang{90}, +\ang{90}]$ \\
$d_r$ [pc] & uniform & $[10, 400]$ \\
$r$ [pc]   & uniform & $[10, 400]$ \\
$v_\perp$ [km\,s$^{-1}$] & normal & $(\mu, \sigma) = (20, 10)$ \\
  \hline
  
\end{tabular}
\end{table}

\begin{table}
\caption{\label{table:C2} Posterior mean and standard deviations for the different models and data sets. The values for $|\vec{v}|$, $l$ and $b$ in the constant flow models were calculated from the posterior samples of $v_x$, $v_y$, and $v_z$.}

\centering
\begin{tabular}{lr@{}l}
parameter & post. mean \,&$\pm$\, post. std.\\
\hline\hline
\textbf{Local -- UV, con. flow model}  \\
$v_x$ &  $-$12.61 \, &$\pm$\,  0.19 \,$\mathrm{km\,s^{-1}}$  \\   
$v_y$ &  5.76 \, &$\pm$\,  0.21 \,$\mathrm{km\,s^{-1}}$  \\
$v_z$ &  1.66 \, &$\pm$\,  0.22 \,$\mathrm{km\,s^{-1}}$  \\
$|\vec{v}|$ & 13.97 \, &$\pm$\,  0.21 \,$\mathrm{km\,s^{-1}}$  \\
$l$ & $-$24.56$^{\circ}$ \, &$\pm$\, 0.84$^{\circ}$ \\   
$b$ & $-$6.81$^{\circ}$ \, &$\pm$\,  0.44$^{\circ}$ \\
  \hline\hline  
\textbf{US -- Ca~\textsc{ii}, con. flow model}  \\
$v_x$ &  $-$11.04 \, &$\pm$\,  2.45 \,$\mathrm{km\,s^{-1}}$  \\   
$v_y$ &  15.48 \, &$\pm$\,  4.32 \,$\mathrm{km\,s^{-1}}$  \\
$v_z$ &  $-$5.98 \, &$\pm$\,  6.10 \,$\mathrm{km\,s^{-1}}$  \\
$|\vec{v}|$ & 21.20 \, &$\pm$\,  3.16 \,$\mathrm{km\,s^{-1}}$  \\
$l$ & $-$53.39$^{\circ}$ \, &$\pm$\, 11.97$^{\circ}$ \\   
$b$ & $+$16.65$^{\circ}$ \, &$\pm$\,  8.16$^{\circ}$ \\
  \hline\hline  
\textbf{US -- Ca~\textsc{ii}, thin shell model}  \\
$v_\perp$ & 18.76 \, &$\pm$\,  2.62 \,$\mathrm{km\,s^{-1}}$  \\ 
$R$ & 241.49 \, &$\pm$\,  92.19 \,$\mathrm{pc}$  \\  
$r$ & 264.83 \, &$\pm$\,  97.00 \,$\mathrm{pc}$  \\   
$l$ & $-$41.27$^{\circ}$ \, &$\pm$\, 12.91$^{\circ}$ \\   
$b$ & $+$17.34$^{\circ}$ \, &$\pm$\, 13.72$^{\circ}$ \\
\hline
  
\end{tabular}
\end{table}

In Table~\ref{table:C2}, the mean and the standard deviations of the respective marginal posterior distributions for the model parameters are summarised for all three cases. For the constant flow models (Fig.~\ref{fig:sky}), we also show the posterior mean and standard deviations of the absolute flow velocity $|\vec{v}|$ and the angular origin of the flow $(l, b)$, all calculated from posterior samples. It should be noted that the origin of the local flow probed in the UV is located in the upwind direction.
The model constrained by the \citet{2008ApJ...673..283R} data shows much smaller uncertainties, which can be attributed to the larger data set and the full sky coverage.
Both the origin and the velocity amplitude of the flow is notably different when comparing the ISM traced by Ca~\textsc{ii} and the UV features studied by \citet{2008ApJ...673..283R}, specifically Mg~\textsc{ii} and Fe~\textsc{ii}. 
While the difference in the velocity amplitudes is statistically significant, the high uncertainty in the determined coordinates of the origin of the Ca~\textsc{ii} flow makes a definitive statement impossible.

\begin{figure*}
    \centering
    \includegraphics{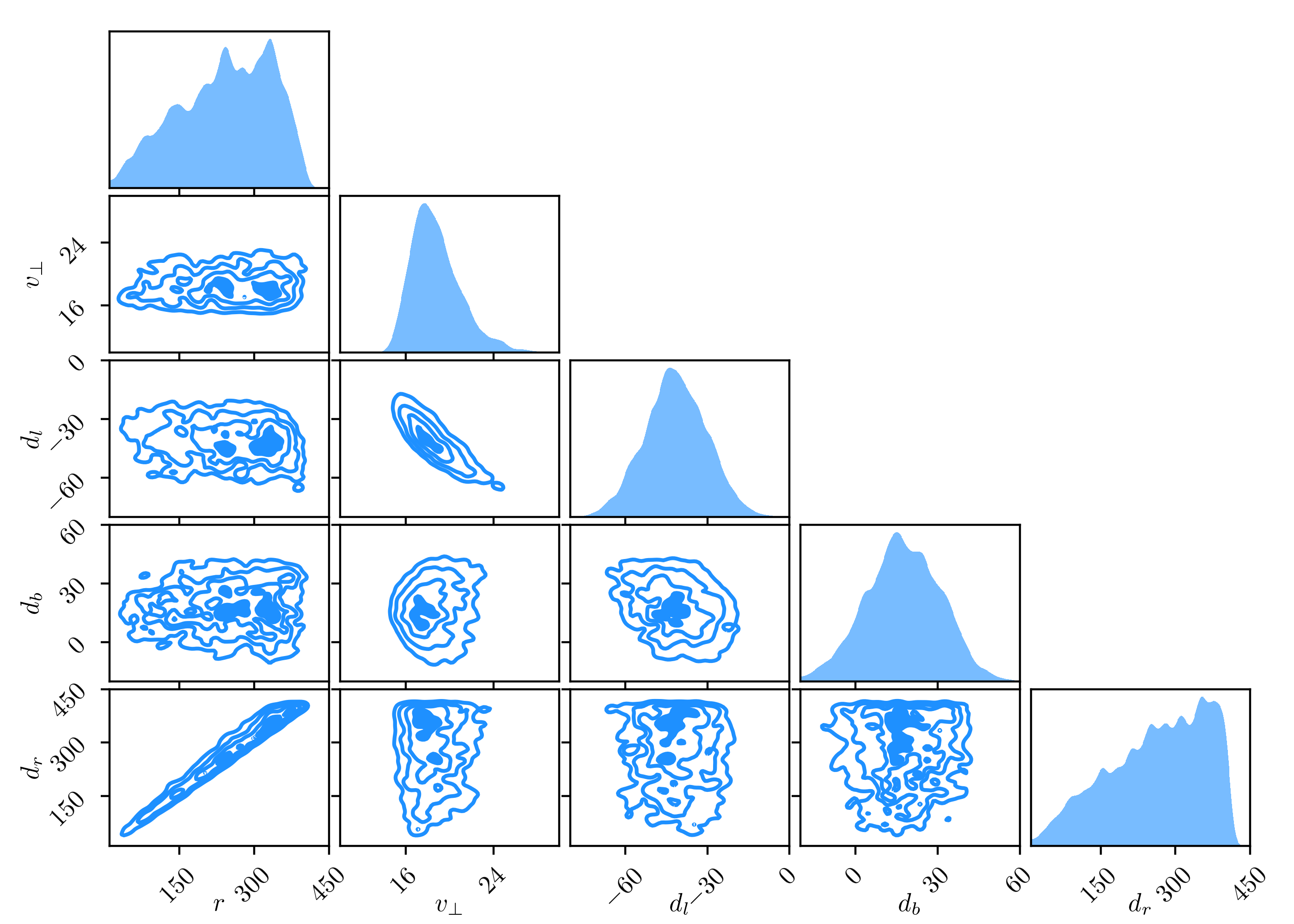}
    \caption{Corner plot illustrating the marginal posteriors of the thin shell model constrained by the Ca~\textsc{ii} data. The one-to-one correlation between the distance parameter ($d_r$) and the radius parameter ($r$) is clearly seen in the bottom left corner. This sub-plot suggests that a single-component spherical flow model cannot explain the observed data and that the model locally approaches (in the limit) a constant flow.}
    \label{fig:C}
\end{figure*}

We note that some of these distributions are significantly skewed, implying that these numbers have to be interpreted with care.
The marginal posteriors of the Ca~\textsc{ii} sky position (thin shell case) are shown in Fig. \ref{fig:C}. 
The comparison of the constant flow models reveals notable differences in both the velocity amplitude and the flow origin. 
The latter plot reveals that the radius and distance of the sphere are strongly correlated, with $R$ being only slightly smaller than $r$ in most cases and the mode of the $R$ posterior close to  the maximum value allowed by the prior (400~pc).
This indicates that the fit prefers the sphere to be as flat as possible, and hence puts its origin far beyond Sco-Cen, rendering these results unlikely, as an origin beyond Sco-Cen would imply a past interaction of the flow with Sco-Cen, making the geometry of a perfect sphere extremely unlikely.
We hence prefer the constant flow model as the best explanation of the observed flow, with the caveat that significant residuals remain in the case of Ca~\textsc{ii}. These results suggest that a single feedback source or event is unlikely to explain the available observed data.  

In summary, we confirm that the flows probed by Ca~\textsc{ii} and UV features originate from Sco-Cen, specifically from regions somewhere between Lupus and Chamaeleon clouds, towards the general location of the Sco-Cen 15~Myr population \cite{2023A&A...678A..71R}. The two identified origins coincide only within 2~$\sigma$ uncertainties, with the posterior based on the calcium spectra covering a much larger area in the sky. It is worth noting that the local flow appears to be slower than the flow identified in Ca~\textsc{ii}.

\section{Connecting calcium and hydrogen kinematics}\label{section:5}

Under the right conditions, H~\textsc{i} can be a very useful tracer of an outflow. Since the ISM density is expected to decrease as a function of the distance from Sco-Cen (origin of the flow), one might also expect a gradual decline of the intensity of the 21-cm line. 
We noticed the presence of two clearly distinguishable structures towards the Sco-Cen massive stars in the \textit{HI4PI} data. 
\begin{enumerate}
    \item We note the presence of an H~\textsc{i} elongated cloud visible at velocities between $-18$~km\,s$^{-1}$ and $-10$~km\,s$^{-1}$. It is about $20^{\circ}$ long, extending from $(l,b) = (357.6^\circ, +32.8^\circ)$ southwards to $(l,b)=(346.4^\circ, +16.3^\circ)$. \citet{1970A&A.....5..135S} first identified this cloud and estimated its distance to be 170 pc. \citet{1986A&A...164..274C} later showed that the cloud is a part of the gas expansion in this region, likely driven by stellar winds.
    \item An H~\textsc{i} ring with a radius of $\sim 9^{\circ}$ at velocities as low as $-30$~km\,s$^{-1}$, and as high as $0$~km\,s$^{-1}$, centred close to the massive stars $\beta$~Sco and $\delta$~Sco. 
    \end{enumerate}
    
\begin{figure}
 \includegraphics[width=\columnwidth]{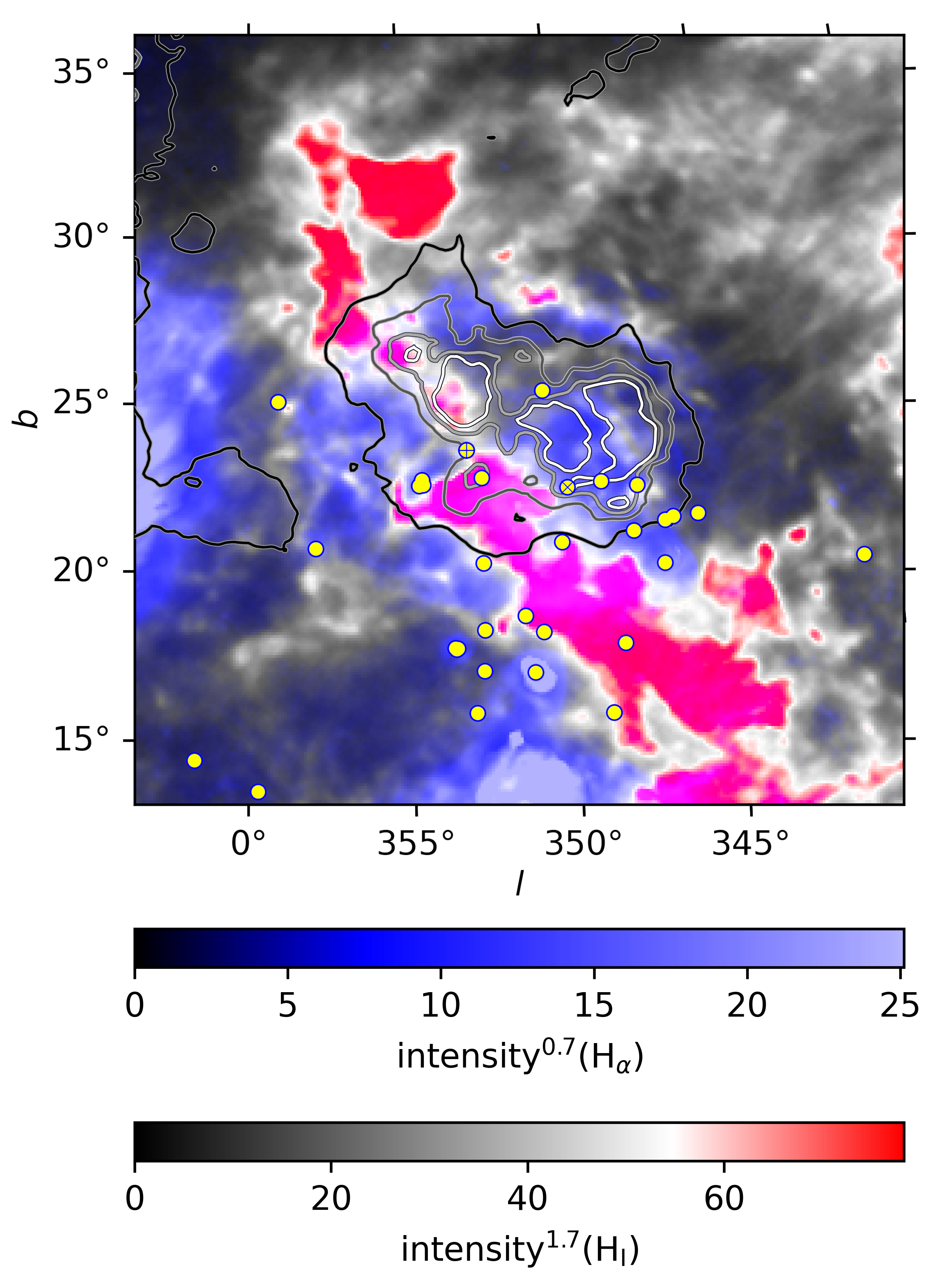}
 \caption{Map of the region around the studied cloud constructed by overlapping the maps in H~\textsc{i} \citep[][$-13$~km\,s$^{-1}$]{2016A&A...594A.116H} and H~\textsc{ii} \citep[][]{2003ApJS..146..407F}. The default map intensities were re-scaled with the use of a power law (see the colour bars) for a better contrast between the images. The stars used to analyse the spectra of Ca~\textsc{ii} are displayed as yellow circles, where $\beta$~Sco and $\delta$~Sco are highlighted by using a plus and a cross symbol, respectively. Contours of the dust map from \citet{GordianDustMap} at the distance of 107~pc are also presented (from black to white: $A_V = 9$, 16, 23, and 30~mmag\,pc$^{-1}$). To view the same image in different colour-maps, see: \href{https://github.com/mpiecka/Sco-Cen-outflow-H-I-filament-sky-map}{https://github.com/mpiecka/Sco-Cen-outflow-H-I-filament-sky-map}.}
 \label{fig:D1}
\end{figure}

The H~\textsc{i} emission projected onto the sky and overlaid with the H~\textsc{ii} emission \citep{2003ApJS..146..407F} can be seen in Fig.~\ref{fig:D1} (black-white-red and black-blue colour-maps, respectively). The elongated filamentary cloud and the ring-like structure can be clearly identified in H~\textsc{i}. Furthermore, we note that the H~\textsc{i} ring appears to enclose the prominent H~\textsc{ii} region Sharpless~7, or Sh2-7, which is being primarily ionised by $\delta$~Sco. 
Given the narrow edge of the ring, its large extension in velocities, and it being centred close to $\beta$~Sco and $\delta$~Sco, we suggest a possible connection to the stellar feedback. We are unable to identify a similar connection for the elongated cloud. The ionised regions in Fig.~\ref{fig:D1} located to the east and to the south are related to $\zeta$~Oph and $\tau$~Sco, respectively.

Using the 3D dust map from \citet{GordianDustMap}, we are able to identify a dust feature at the coordinates of the H~\textsc{i} cloud $(l,b)=(353.8^\circ, +25.5^\circ)$. We determine the distance towards this cloud to be between 95~pc and 120~pc. The nominal distance is identified at the distance channel showing the highest extinction, specifically at 107~pc. Since the determined distance to the closest star ($\beta$~Sco) is about 125~pc, the hydrogen outflow material must extend to at least 20~pc away from the source of the flow. We note that the distances to the most important outflow drivers in this region ($\delta$~Sco, $\beta$~Sco) based on the properties of their hosting clusters \citep{2023A&A...677A..59R} suggest that they are located further away from the cloud, between 140 and 155 pc away from the Sun.

\begin{figure}
 \includegraphics[width=\columnwidth]{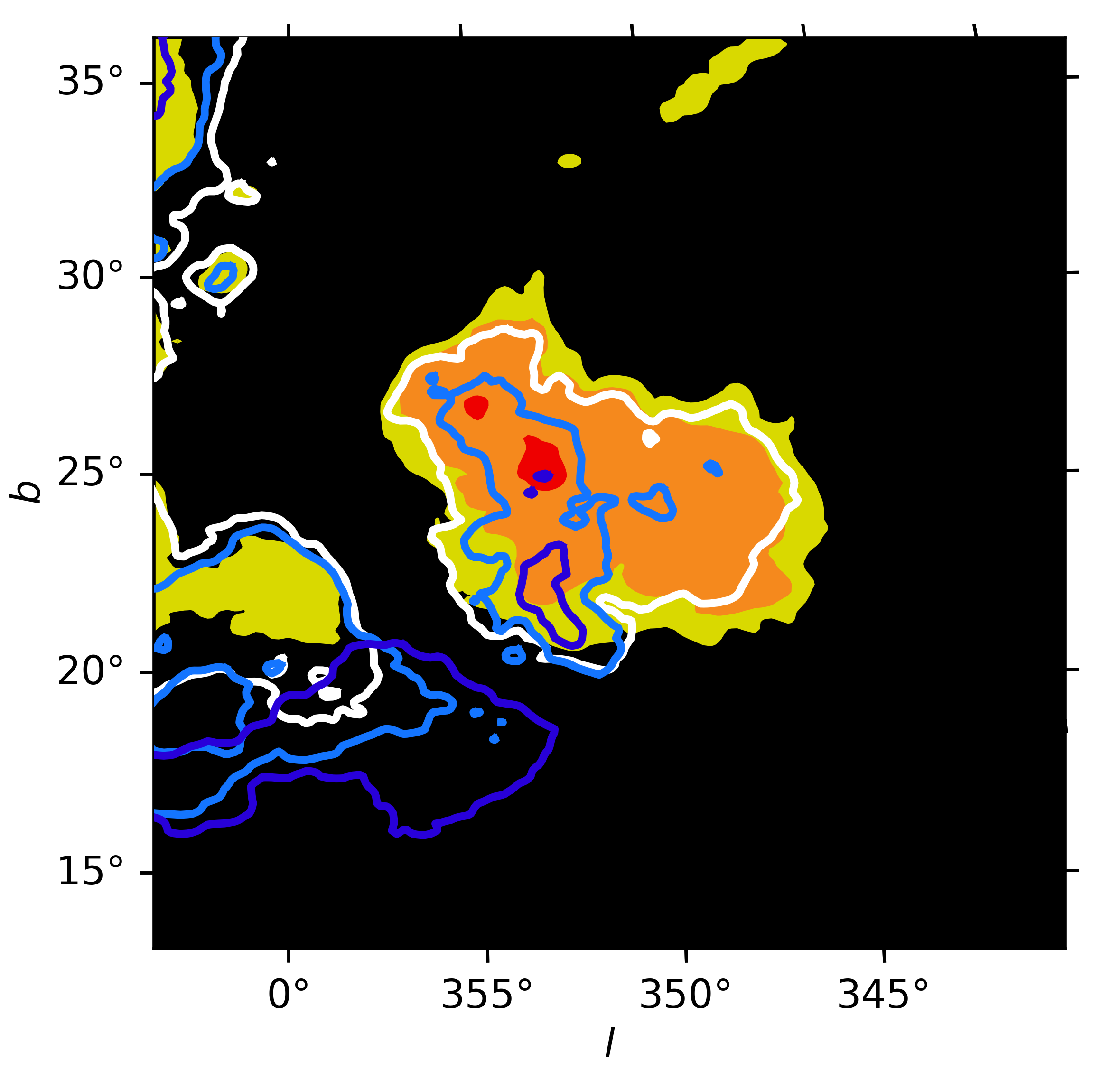}
 \caption{Slices of the dust map from \citet{GordianDustMap} at six different distances in the area of the H~\textsc{i} cloud (same as Fig.~\ref{fig:D1}). The contours represent $A_V = 9$~mmag\,pc$^{-1}$. The foreground regions at 95~pc, 100~pc, and 105~pc are displayed as filled contours in red, orange, and yellow, respectively. The background regions at 110~pc, 115~pc and 120~pc are presented as white, blue, and dark blue contour lines, respectively.}
 \label{fig:D2}
\end{figure}

The contours of the dust cloud are displayed in Fig.~\ref{fig:D1} -- the eastern side of the dust cloud is located on top of the northern half of the H~\textsc{i} cloud (close to its centre). On the other hand, the western portion of the dust cloud has a shape resembling Sh2-7. The contours of the dust map ($A_V = 9$~mmag) at various distances are displayed in Fig.~\ref{fig:D2}, indicating that the western part of the dust cloud is less extended in the distance than its eastern part.

The positional and the kinematic structures of the H~\textsc{i} cloud and the ring display variations that we presently ignore. A quick look at various velocity bins (from $-30$ to $+10$~km\,s$^{-1}$) of the H~\textsc{i} map reveals sudden changes ($\sim 1$~km\,s$^{-1}$, $\sim 1^{\circ}$) in the morphology of the structures in the sky. Additional complexity is introduced when one starts to look at very high absolute value of the RVs \citep[see for example][]{2018A&A...617A.101R}. Furthermore, the bright endpoints of the H~\textsc{i} cloud remain obvious at negative velocities as high as $-30$~km\,s$^{-1}$, in contrast with the rest of the elongated cloud that appears to span $\approx 8$~km\,s$^{-1}$.

\begin{figure}
 \includegraphics[width=\columnwidth]{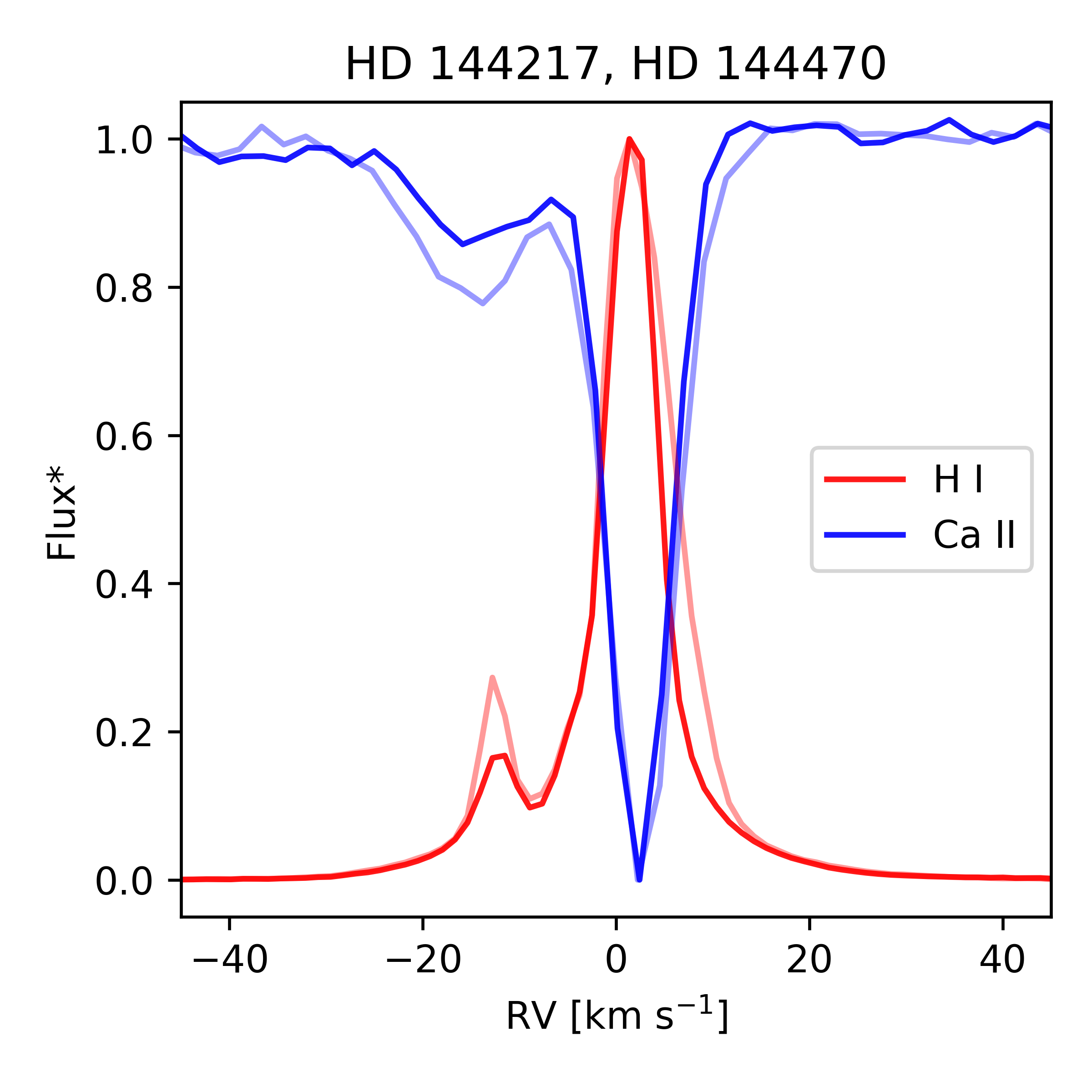}
 \caption{Comparison of H~\textsc{i} and Ca~\textsc{ii} spectra towards $\beta^1$~Sco (HD~144217, darker colours) and $\omega$~Sco (HD~144470). Both of these stars are located near the centre of the studied H~\textsc{i} cloud. The fluxes were stretched so that both the absorption and the emission features are bounded between 0 and 1.}
 \label{fig:D3}
\end{figure}

To investigate a potential transition from the hydrogen flow to the previously described calcium flow, we analysed H~\textsc{i} spectra from the \textit{HI4PI} data cube towards our calcium-probing targets listed in Table~\ref{table:A}. A comparison of the H~\textsc{i} and Ca~\textsc{ii} line profiles (Fig.~\ref{fig:D3}) reveals a striking resemblance in most cases, particularly towards the sources probing the H~\textsc{i} cloud. This similarity extends to the hydrogen line's overall shape and potential sub-structure, suggesting a strong connection between the two flows. This connection is critical in establishing the distance to the flow traced by Ca~\textsc{ii}.
We also note a possible existence of small systematic differences in the velocities of calcium and hydrogen -- in most lines-of-sight, the differences are smaller than $5$~km\,s$^{-1}$ (between 2 and 3~km\,s$^{-1}$ in Fig.~\ref{fig:D3}).

\section{Comparison with literature}\label{section:6}

There are many excellent published papers addressing the same topic as this paper. Although we do not intend to provide a comprehensive review \citep[we refer the reader to see][]{1995SSRv...72..499F,2011ARA&A..49..237F}, several studies closely related to our analysis warrant special attention. Below, we compare our results with those from five other key studies.

\subsection{Crawford (1991)}\label{section:6.1}

This paper was already mentioned in Section~\ref{section:1}. While the most negative velocity component of Ca~\textsc{ii} and Na~\textsc{i} was found to be at around $-20$~km\,s$^{-1}$, the outflow models analysed by \citet{1991A&A...247..183C} suggest a much smaller flow velocity amplitude between 7 and 10~km\,s$^{-1}$. When compared with the local ISM, this would suggest an acceleration of the flow as a function of the distance from the source of the outflow. On the contrary, our analysis suggests a possibility of flow deceleration. The type of the fitted model (an expanding shell and a constant flow) does not significantly change the value of the velocity amplitude, a result achieved in both works.

It should be pointed out that unlike in our work, \citet{1991A&A...247..183C} used stars across the whole Sco-Cen ($80^{\circ} \times 30^{\circ}$) in their analysis. Despite this, our results agree in terms of the origin of the outflow, which seems to be located westwards from the Upper~Sco and above the Galactic disk. The relatively low number (23) of studied lines of sight limits the precision of the flow modelling presented by \citet{1991A&A...247..183C}, while in our cases the limiting factor is the relatively low sky coverage ($30^{\circ} \times 20^{\circ}$). Extending the sky coverage to the whole Sco-Cen (or beyond) and keeping the high number of studied sources per area presented in our work should provide the best possible data set for further investigations of the Sco-Cen outflow when making use of the optical interstellar absorption lines.

\subsection{Frisch (1995) and Frisch et al. (2011)}\label{section:6.2}

\citet{1995SSRv...72..499F} and \citet{2011ARA&A..49..237F} presented outstanding reviews of the past studies focused on the local ISM. While the main focus of these works was put on the $<20$~pc distant local clouds, this topic connects to our work via kinematics. As was already mentioned, the RVs obtained for the local ISM put the origin of the local ISM to a region somewhere in Sco-Cen. Since an outflow can be connected to both, the region very close to the Sun and to the regions within Sco-Cen, it is clear that there must be either continuity of the outflow or that the outflow parts are the results of multiple flow-driving events.

Our work confirms many of the results referenced in the mentioned review papers. For example, both papers show that the observed RVs (in LSR) follow a cosine law when plotted as a function of the Galactic longitude or the angular distance from the source of the flow. This is true for both, Ca~\textsc{ii} \citep[Figure~6,][]{1995SSRv...72..499F} and the local ISM probing UV features \citep[Figure~3 in both,][]{2002ApJ...574..834F,2011ARA&A..49..237F}. This behaviour is expected from a flow that can be best described by using a constant vector field, as was discussed in \citet{1991A&A...247..183C}.

Furthermore, \citet{1995SSRv...72..499F} discussed the neutral hydrogen kinematics. In this case, we are able to provide a more precise distance measurement to the H~\textsc{i} cloud originally presented by \citet{1970A&A.....5..135S}. In the future works, this distance determination should help to better understand the processes that drive the gas flow, especially when it comes to the interaction of the Sco-Cen outflow driving force with the relatively low-density dust clouds.

\subsection{Krause et al. (2018)}\label{section:6.3}

\citet{2018A&A...619A.120K} used observations and theory to try and characterise the established Sco-Cen outflow. The authors compared the H~\textsc{i} velocities with the Na~\textsc{i} spectra combined with the stellar parallaxes, similar to our analysis based on Ca~\textsc{ii}. Unlike us, the authors' primary use of the optical ISM line was to constrain gas distances. This was done by associating components of the H~\textsc{i} line profile with the components observed within the Na~\textsc{i} profile, yielding an upper distance limit for each line of sight. The region probed by \citet{2018A&A...619A.120K} includes the H~\textsc{i} cloud and ring analysed in this work.

It is precisely the multiple blue-shifted components that draw our attention. Presented in their Figure~5 (right panel), \citet{2018A&A...619A.120K} suggest that these components are "co-spatial but separated in the velocity space". Comparing results from both works, it seems that their blue-shifted (tube-like) structures might be related to the gas ring identified in this work. However, the prominent filamentary cloud appears to be missing in their results -- due to its size, it should appear as a large (in dimensions similar to their Upper~Sco loop) structure at around $-13$~km\,s$^{-1}$.

As was mentioned above, the existence of a local ISM that is kinematically connected to the gas flow observed near/within Upper~Sco points towards a continuity of the Sco-Cen outflow. Re-examining the results of our work, we find no clear evidence that would support (or contradict) the claim of a co-spatial structure mentioned by \citet{2018A&A...619A.120K}. The discussion regarding the hydrogen ring presented in Section~\ref{section:5} might hint at this possibility but we prefer to avoid speculating about this topic. Further research is certainly required to provide additional information about this property of the flow.

In general, we find an agreement with the results obtained by \citet{2018A&A...619A.120K}. Their main point is the interaction of two super-bubbles, with the interface between them containing denser gas (and dust) structure. This interface seems to coincide with the intervening material located around 110~pc away from the Sun, visible in front of $\zeta$~Oph and $\beta$~Sco in our Fig.~\ref{fig:stage}. The complex kinematic properties of this gas component observed in H~\textsc{i} suggest an interaction between the outflow driving force at the interface material. The sub-structure identified in Ca~\textsc{ii} line profile in some lines of sight might be partially related to the same process, but this needs to be confirmed with higher resolution spectra (obtained, for example, using HARPS).

\subsection{Linsky et al. (2022)}\label{section:6.4}

\citet{2022AJ....164..106L} give an update on the information about the local clouds that are part of the local flow, building up primarily on the results presented in \citet{2008ApJ...673..283R}. The updated data show hints of variations in the velocity dispersion and temperatures of the local ISM. If proven to be a statistically significant result, it would be interesting to identify the processes leading to these conditions -- could these variations be a general property of the Sco-Cen outflow?

\section{Discussion and conclusions}\label{section:7}

\begin{figure*}
 \centering
 \includegraphics[width=2\columnwidth]{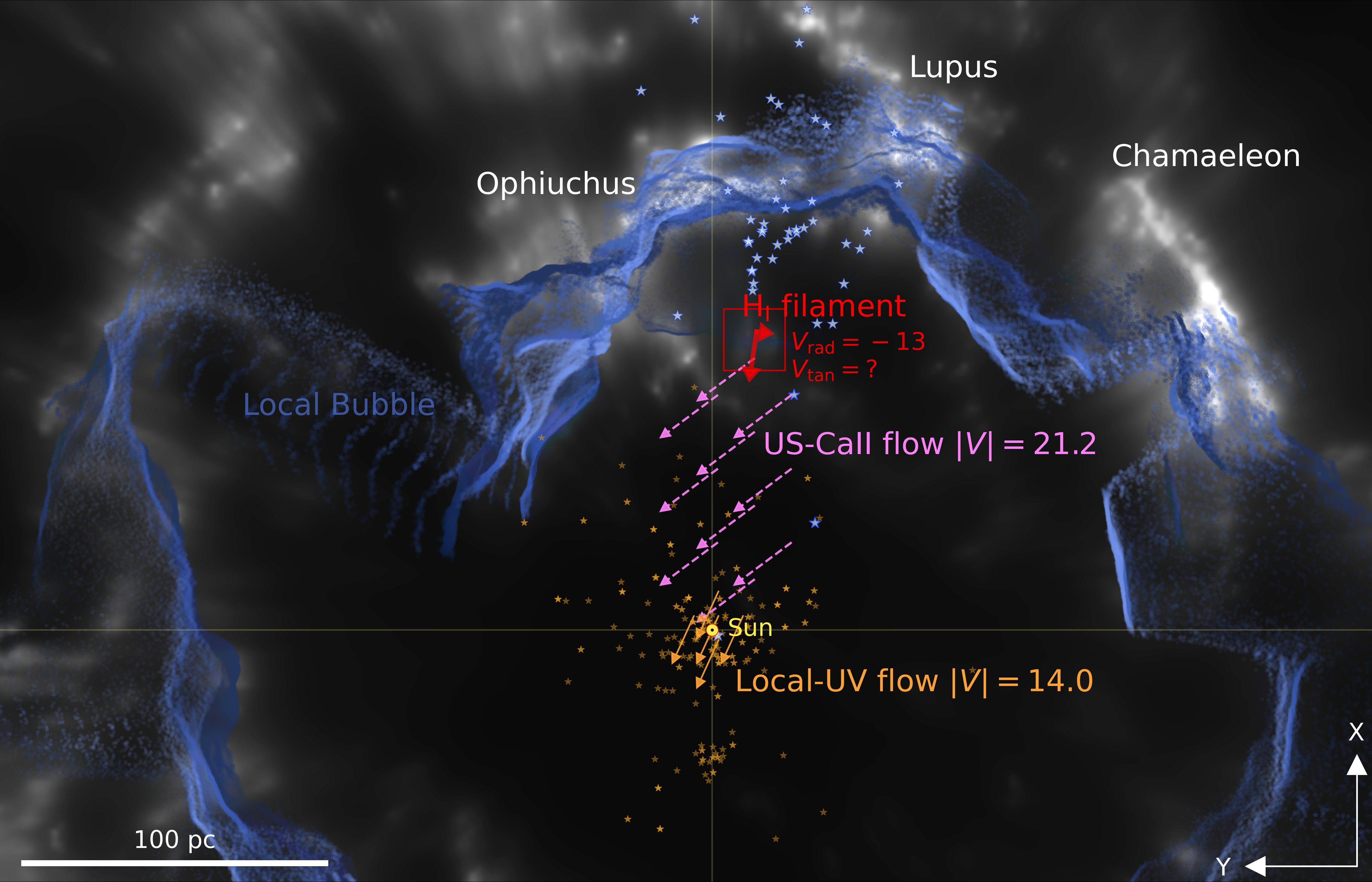}
 \caption{Map of the region between the Sun and Sco-Cen. The grey-shaded areas represent the dust map from \citet{GordianDustMap} projected onto the $XY$-plane, while the blue-shaded areas (brighter towards the Galactic plane) display the Local Bubble map \citep[][$-50 < Z < 50$ pc]{TheoLocalBubble} based on the same dust map. The red rectangle highlights the position of an H~\textsc{i} cloud. The smaller orange stars show the positions of the objects from \citet{2008ApJ...673..283R}, where brighter data points represent measurements with multiple velocity components. The larger blue stars indicate the positions $\left(d = \varpi ^{-1}\right)$ of our calcium-probing (mostly B-type) stars. The two arrows in the bottom right corner show the directions towards the Galactic centre (up) and the Galactic rotation (left). The bar in the bottom left corner (length of 100~pc) serves as a measure of the distance scale in the map. Coloured arrows display the direction of the flows probed by different data, with the scatter in the arrows highlighting the uncertainty in the distance towards the corresponding flow component -- any of these arrows can be picked to represent the flow vector. The displayed velocity values are in km\,s$^{-1}$. In the case of the cloud, we have no knowledge of its tangential motion. We note that the Ca~\textsc{ii} flow is based on spectra with unresolved kinematic sub-structure.}
 \label{fig:act}
\end{figure*}

In this study, we connected observational evidence from nearby stars and those in the Sco-Cen association, indicating the presence of a general interstellar outflow originating from the OB association. We utilised two different probes of the ISM: the Ca~\textsc{ii} line profiles obtained from FEROS spectra and the radial velocities (RVs) derived from the UV (Mg~\textsc{ii}, Fe~\textsc{ii}) spectra by \citet{2008ApJ...673..283R}. By fitting a constant flow model (Fig.~\ref{fig:sky}), we confirm that both flows appear to originate from a region between the Lupus and the Chamaeleon clouds, or towards the direction of the major star formation event in Sco-Cen, where most of the massive stars and largest clusters in the association formed $\approx 15$ Myr ago \citep{2023A&A...677A..59R}. We also found significant differences in the exact origin and flow velocity amplitudes, suggesting the existence of distinct kinematic (and possibly spatial) components of the Sco-Cen outflow.

Within the flow, we identified a previously understudied H~\textsc{i} cloud with the concurrent detection in H~\textsc{i} and Ca~\textsc{ii} at about the same radial velocity.
However, small systematic differences ($<5$~km\,s$^{-1}$) in the velocities of Ca~\textsc{ii} and H~\textsc{i} are noted.
This suggests a more intricate flow structure than previously assumed.
We speculate that the calcium flow may be interacting with the H~\textsc{i} cloud.

A representation of the Sco-Cen outflow projected onto the Galactic plane is displayed in Fig.~\ref{fig:act} -- the dust feature that we identify with the H~\textsc{i} (discussed in Section~\ref{section:5}) is highlighted. We find supporting evidence that the Sco-Cen outflow is an ongoing process and can be linked to the massive stars in Sco-Cen. The local flow (as traced in the UV absorption lines) and the flow probed by Ca~\textsc{ii}, at a more uncertain distance, together with the properties of the H~\textsc{i} cloud all indicate the existence of an ISM reaching from Sco-Cen to the current position of the Sun within the Local Bubble. However, there are many questions that remain to be answered, including:

\begin{itemize}
    \item 
    \textit{Why is the observed local flow so uniform?} Our modelling yielded the surprising result that the local flow is extremely uniform and that a single spherical-flow model cannot adequately explain the global observations. This points to a complexity that needs to be further explored.
    \item \textit{Do the observed flows originate from a common region?} We cannot rule in or out a common flow origin based on the calcium and the local UV absorption measurements. However, using higher-resolution spectra ($\textrm{R}>100\,000$) together with additional calcium observations at higher angular separation from Upper~Sco would significantly constrain the posterior distribution of the calcium flow model and provide an answer to this question.
    \item \textit{Has the flow been primarily shaped by supernovae or stellar winds and radiation?} Presently, we cannot determine which drivers dominate the observed flows. The driving force behind the H~\textsc{i} cloud may differ from the one that shaped the motion of the local ISM.
\end{itemize}

Answering these questions requires further research. For example, we should be able to gain an additional insight into the structure of the outflow by including a larger number of observations such as the one based on HARPS and presented in Fig.~\ref{fig:B}. This should lead to uncovering of the kinematic sub-structure of the outflow, extending the number of components beyond the two presented in this Paper.

The Sco-Cen flow feeds the Local Bubble \citep{Breitschwerdt2016-le,Zucker2022-hw}. This work takes a critical step towards exploring the closest large-scale ISM outflow. It confirms that the flow found in the very local (d $<$ 30 pc) ISM is part of a larger flow from Sco-Cen. Our analysis raises more questions than answers but calls attention to an important and exciting aspect of the ISM in the solar neighbourhood, the one the Sun is currently crossing. The presence of supernova radioisotopes on Earth \citep{2016Natur.532...69W,Koll2019-eg}, a strong argument in favour of a large-scale ISM outflow from Sco-Cen, is a strong motivation for further studies of this flow. 

Finally, large 100-pc scale outflows, like the one studied here, are the natural consequence of massive star formation events. The common occurrence of feedback-driven bubbles \citep[e.g.,][]{JWST}, suggests these outflows are an important component of the ISM in spiral galaxies. Kinematic studies of other local ISM flows powered by massive star formation can provide important information about the evolution of the galactic environment.

\begin{acknowledgements}
Co-funded by the European Union (ERC, ISM-FLOW, 101055318). Views and opinions expressed are, however, those of the author(s) only and do not necessarily reflect those of the European Union or the European Research Council. Neither the European Union nor the granting authority can be held responsible for them.
This paper made use of data obtained from the ESO Science Archive Facility, specifically from the following ESO programmes: 0106.A-9009(A), 073.C-0337(A), 076.C-0164(A), 077.C-0575(A), 078.C-0403(A), 081.C-2003(A), 082.B-0610(A), 082.D-0061(A), 083.D-0034(A), 086.D-0236(A), 086.D-0449(A), 087.A-9005(A), 089.D-0153(A), 090.D-0358(A), 091.C-0713(A), 094.A-9012(A), 097.A-9024(A), 099.A-9029(A), 179.C-0197(A), 179.C-0197(C), 183.C-0972(A), 60.A-9036(A), and 60.A-9700(G).

This work made use of data from the European Space Agency (ESA) mission {\it Gaia} (\url{https://www.cosmos.esa.int/gaia}), 
processed by the {\it Gaia} Data Processing and Analysis Consortium (DPAC, \url{https://www.cosmos.esa.int/web/gaia/dpac/consortium}). 
Funding for the DPAC has been provided by national institutions, in particular the institutions participating in the {\it Gaia} 
Multilateral Agreement.

The following Python libraries were used in this work: \texttt{numpy} \citep{numpy}, \texttt{scipy} \citep{scipy}, \texttt{astropy} \citep{astropy3}, \texttt{dynesty} \citep{2020MNRAS.493.3132S, sergey_koposov_2023_8408702}, \texttt{cmasher} \citep{2020JOSS....5.2004V}, and \texttt{matplotlib} \citep{matplotlib}.

The authors are grateful to M.~Kajan for his valuable suggestions regarding the treatment of stellar spectra. Furthermore, we express our gratitude to C.~Zucker and J.~Linsky for the discussion about the local flow.

\end{acknowledgements}

\bibliographystyle{aa} 
\bibliography{aanda}

\begin{appendix}

\section{Determined radial velocities}\label{section:A}

\begin{figure}
 \includegraphics[width=\columnwidth]{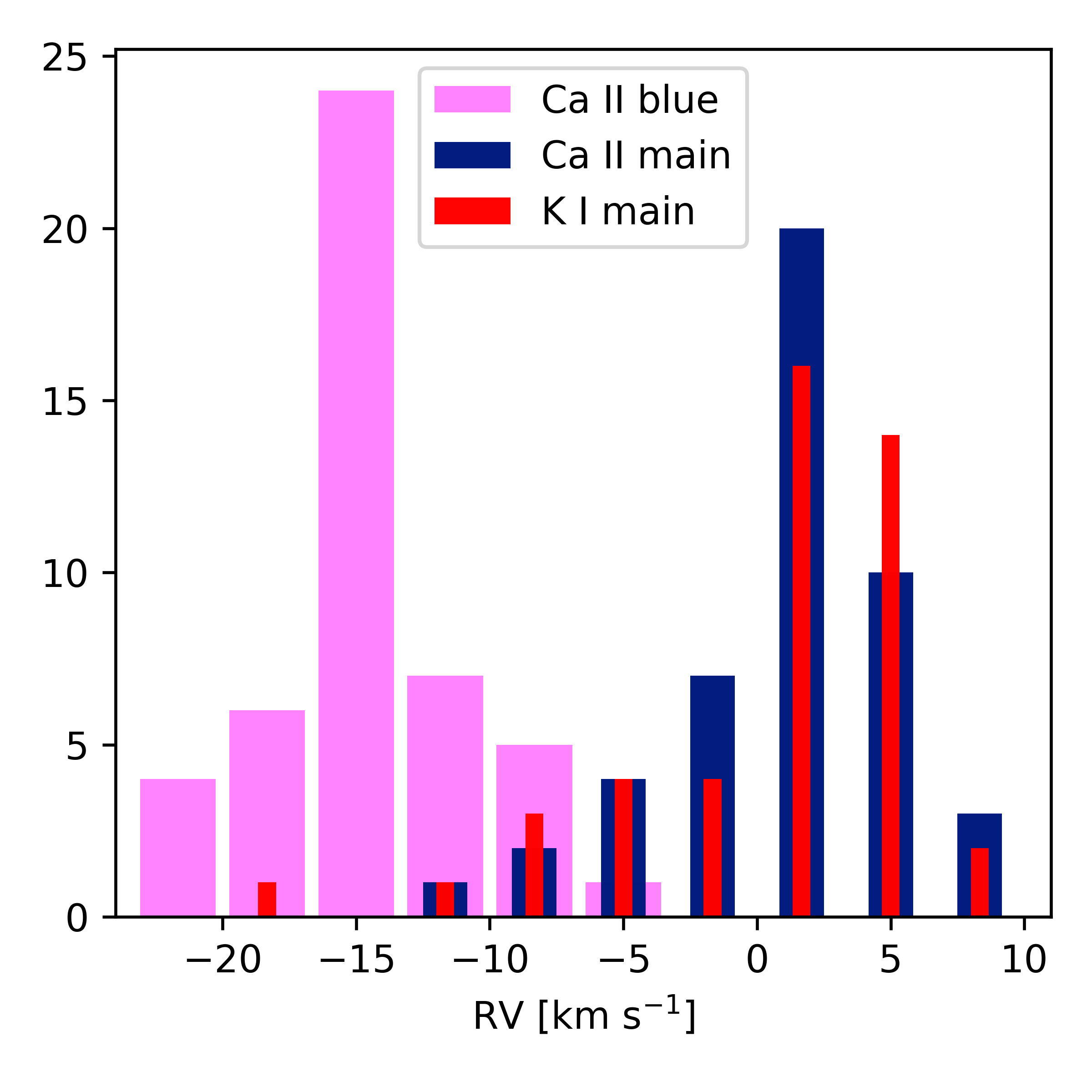}
 \caption{Histogram of the measured interstellar radial velocities for our 47~targets. We note that at most one of the lines of sight shows a blue-shifted component in the spectrum of K~\textsc{i}.}
 \label{fig:A}
\end{figure}

The results of our Ca~\textsc{ii} line-fitting procedure are presented in Table~\ref{table:A} and in Fig.~\ref{fig:A}. The values were obtained by making use of a bootstrapping method with $N=10^4$ repetitions. The derived values correspond to the 50th percentiles of the determined distributions. The numerical errors, $\epsilon$, were calculated as half of the difference between the 84th and the 16th percentiles. The reported errors $\left(\Delta = \sqrt{\epsilon^2 + \sigma^2}\right)$ include a minimum error limit, $\sigma$, which was evaluated as the distance between the neighbouring points of a spectrum at a given wavelength (and thus represents the spectral resolution). We find that these errors tend to be more accurate than those derived by using a single-iteration \texttt{scipy.optimize.curve\_fit}. We note that $\Delta \approx \sigma$ when the noise is low enough for the feature to be clearly distinguishable.

\begin{table*}
\caption{List of the analysed OB stars around Upper~Sco, including the results of the spectral fitting for Ca~\textsc{ii} and K~\textsc{i}. Asterisk denotes astrometric data obtained from Hipparcos instead of Gaia (used in Figs.~\ref{fig:stage} and \ref{fig:act}). The second column displays ESO program IDs for the stellar spectra used to obtain information about the studied interstellar lines. Indices for the equivalent widths (EW) and the radial velocities (RV) are used to label the analysed features: the blue-shifted Ca~\textsc{ii} component (1), the main Ca~\textsc{ii} component (2), and the main K~\textsc{i} component (3). The errors are typically dominated by the imposed minimum value.}\label{table:A}
\small
\centering
\begin{tabular}{ll|rrr|rrr|rrr}
\hline\hline
Star & ESO program ID & $\textrm{EW}_1$ & $\textrm{RV}_1$ & $\Delta\textrm{RV}_1$ & $\textrm{EW}_2$ & $\textrm{RV}_2$ & $\Delta\textrm{RV}_2$ & $\textrm{EW}_3$ & $\textrm{RV}_3$ & $\Delta\textrm{RV}_3$  \\
 & & [m\AA] & [km\,s$^{-1}$] & [km\,s$^{-1}$] & [m\AA] & [km\,s$^{-1}$] & [km\,s$^{-1}$] & [m\AA] & [km\,s$^{-1}$] & [km\,s$^{-1}$] \\ 
  \hline
  
HD116658* & 073.C-0337(A) & 2.1 & -7.2 & 2.3 & ... & ... & ... & ... & ... & ... \\
HD133529 & 179.C-0197(A) & 13.8 & -13.7 & 2.2 & 49.9 & 2.6 & 2.2 & 50.2 & 2.7 & 1.2 \\
HD139365* & 097.A-9024(A) & 2.1 & -23.0 & 2.6 & 2.8 & -12.4 & 3.5 & 6.7 & -8.5 & 1.9 \\ 
HD140008* & 091.C-0713(A) & 20.1 & -21.8 & 2.4 & 107.7 & 7.5 & 2.2 & 1.2 & 4.5 & 18.9 \\
HD140840 & 082.D-0061(A) & 3.2 & -17.1 & 2.5 & 4.1 & -0.4 & 3.7 & 2.9 & -16.7 & 17.5 \\ 
HD141637 & 086.D-0236(A) & 10.9 & -15.0 & 4.8 & 16.8 & 2.4 & 2.3 & 15.0 & 2.0 & 1.2 \\  
HD142096 & 073.C-0337(A) & 21.7 & -14.8 & 2.2 & 11.5 & 1.0 & 2.2 & 11.7 & 0.3 & 1.2 \\  
HD142114* & 073.C-0337(A) & 5.9 & -18.3 & 2.3 & 8.9 & -4.7 & 2.2 & 11.6 & -6.2 & 1.2 \\ 
HD142184 & 073.C-0337(A) & 21.7 & -13.4 & 2.6 & 15.4 & -1.1 & 2.2 & 28.0 & -2.2 & 1.2 \\
HD142301 & 086.D-0449(A) & 4.7 & -15.9 & 2.4 & 27.2 & 0.0 & 2.2 & 4.3 & -5.3 & 1.2 \\   
HD142705 & 099.A-9029(A) & 8.0 & -20.8 & 2.2 & 29.3 & -6.6 & 2.2 & 32.6 & -6.2 & 1.2 \\
HD142990 & 089.D-0153(A) & 9.1 & -15.0 & 2.3 & 18.0 & -2.4 & 2.3 & 5.9 & -1.9 & 1.6 \\
HD143018* & 097.A-9024(A) & 4.3 & -17.9 & 2.3 & 3.1 & -3.7 & 2.9 & 7.1 & -8.0 & 1.4 \\
HD143118* & 073.C-0337(A) & 4.4 & -18.0 & 2.2 & 9.7 & -3.9 & 2.2 & 0.1 & -7.1 & 24.8 \\
HD143275* & 087.A-9005(A) & 7.8 & -15.3 & 2.2 & 28.2 & 0.0 & 2.2 & 31.3 & -0.2 & 1.2 \\
HD143699* & 097.A-9024(A) & 3.1 & -18.7 & 2.4 & 16.6 & -7.5 & 2.2 & 3.6 & -4.9 & 8.8 \\
HD143927 & 083.D-0034(A) & 3.8 & -15.8 & 4.2 & 22.0 & 6.3 & 2.2 & 30.2 & 7.0 & 1.2 \\
HD144217* & 091.C-0713(A) & 7.8 & -13.9 & 2.2 & 30.9 & 2.4 & 2.2 & 22.6 & 2.9 & 1.2 \\
HD144218 & 073.C-0337(A) & 20.4 & -11.1 & 2.2 & 30.7 & 2.5 & 2.2 & 25.3 & 2.7 & 1.2 \\
HD144294 & 097.A-9024(A) & 10.3 & -15.9 & 2.4 & 2.5 & -2.6 & 5.7 & 2.5 & -11.6 & 16.8 \\
HD144334 & 086.D-0449(A) & 10.9 & -11.7 & 2.2 & 7.7 & 2.2 & 2.2 & 0.6 & 2.2 & 7.4 \\
HD144470* & 090.D-0358(A) & 11.5 & -14.9 & 2.2 & 28.2 & 2.7 & 2.2 & 19.9 & 2.0 & 1.2 \\
HD145102 & 077.C-0575(A) & 24.0 & -3.7 & 2.3 & 16.7 & 6.6 & 2.2 & 8.0 & 5.1 & 1.2 \\
HD145501 & 086.D-0449(A) & 4.5 & -15.2 & 2.2 & 34.1 & 2.1 & 2.2 & 30.1 & 3.1 & 1.2 \\
HD145502* & 073.C-0337(A) & 7.0 & -15.9 & 2.2 & 57.9 & 2.1 & 2.2 & 37.5 & 1.9 & 1.2 \\
HD145554 & 094.A-9012(A) & 7.3 & -15.2 & 2.2 & 41.4 & 2.1 & 2.2 & 34.7 & 2.4 & 1.2 \\
HD145631 & 094.A-9012(A) & 7.5 & -15.7 & 2.2 & 35.6 & 3.2 & 2.2 & 45.5 & 3.4 & 1.2 \\
HD146029 & 179.C-0197(A) & 32.6 & -10.5 & 4.2 & 24.4 & 3.7 & 2.3 & 14.0 & 4.7 & 1.4 \\
HD146254 & 179.C-0197(C) & 5.8 & -9.3 & 5.5 & 69.8 & -0.5 & 2.4 & 42.8 & 4.1 & 1.2 \\
HD146284 & 077.C-0575(A) & 15.9 & -8.9 & 2.2 & 43.0 & 2.7 & 2.2 & 49.6 & 3.5 & 1.2 \\
HD146285 & 083.D-0034(A) & 18.9 & -9.0 & 2.2 & 47.8 & 4.5 & 2.2 & 22.4 & 5.4 & 1.2 \\
HD146606 & 099.A-9029(A) & 5.7 & -16.6 & 2.3 & 2.5 & -0.1 & 5.6 & 3.1 & 3.0 & 1.8 \\
HD147165* & 091.C-0713(A) & 10.7 & -9.0 & 3.3 & 38.9 & 5.1 & 2.2 & 25.7 & 5.2 & 1.2 \\
HD147196 & 081.C-2003(A) & 6.8 & -16.9 & 2.3 & 26.5 & 3.5 & 2.2 & 52.2 & 5.0 & 1.2 \\
HD147683 & 077.C-0575(A) & 24.8 & -10.3 & 2.2 & 82.1 & 8.1 & 2.2 & 93.4 & 8.8 & 1.2 \\
HD147888 & 076.C-0164(A) & 6.9 & -13.9 & 2.4 & 57.6 & 3.3 & 2.2 & 91.2 & 3.5 & 1.2 \\
HD147889 & 081.C-2003(A) & 8.2 & -14.5 & 2.7 & 48.5 & 4.2 & 2.2 & 79.6 & 4.4 & 1.2 \\
HD147932 & 179.C-0197(A) & 6.7 & -14.5 & 2.4 & 56.6 & 3.2 & 2.2 & 81.6 & 3.6 & 1.2 \\
HD147934 & 076.C-0164(A) & 4.8 & -15.0 & 2.5 & 62.5 & 3.6 & 2.2 & 87.9 & 3.7 & 1.2 \\
HD148184 & 081.C-2003(A) & 34.9 & -12.6 & 2.2 & 43.0 & 2.2 & 2.2 & 72.9 & 2.1 & 1.2 \\
HD148605* & 073.C-0337(A) & 7.2 & -14.2 & 2.5 & 15.2 & 3.3 & 2.2 & 7.7 & 2.4 & 1.2 \\
HD149438* & 0106.A-9009(A) & 5.0 & -15.1 & 2.2 & 25.1 & 8.5 & 2.2 & 2.8 & 5.0 & 15.5 \\
HD149757* & 094.A-9012(A) & 12.1 & -13.2 & 2.2 & 36.6 & -0.5 & 2.2 & 65.5 & -0.2 & 1.2 \\
HD152655* & 077.C-0575(A) & 29.9 & -13.4 & 2.2 & 71.6 & 4.5 & 2.2 & 46.3 & 3.0 & 1.2 \\
HD152909 & 077.C-0575(A) & 8.9 & -16.6 & 2.2 & 84.7 & 4.1 & 2.2 & 48.0 & 1.4 & 1.2 \\
HD155503 & 179.C-0197(A) & 23.2 & -10.8 & 2.3 & 36.3 & 1.1 & 2.2 & 29.1 & 1.4 & 1.2 \\
HD158427* & 179.C-0197(C) & 3.7 & -7.4 & 2.4 & ... & ... & ... & ... & ... & ... \\

\hline
\end{tabular}
\end{table*}

\section{Sub-structure in the studied Ca~\textsc{ii} lines}\label{section:B}

\begin{figure}
 \includegraphics[width=\columnwidth]{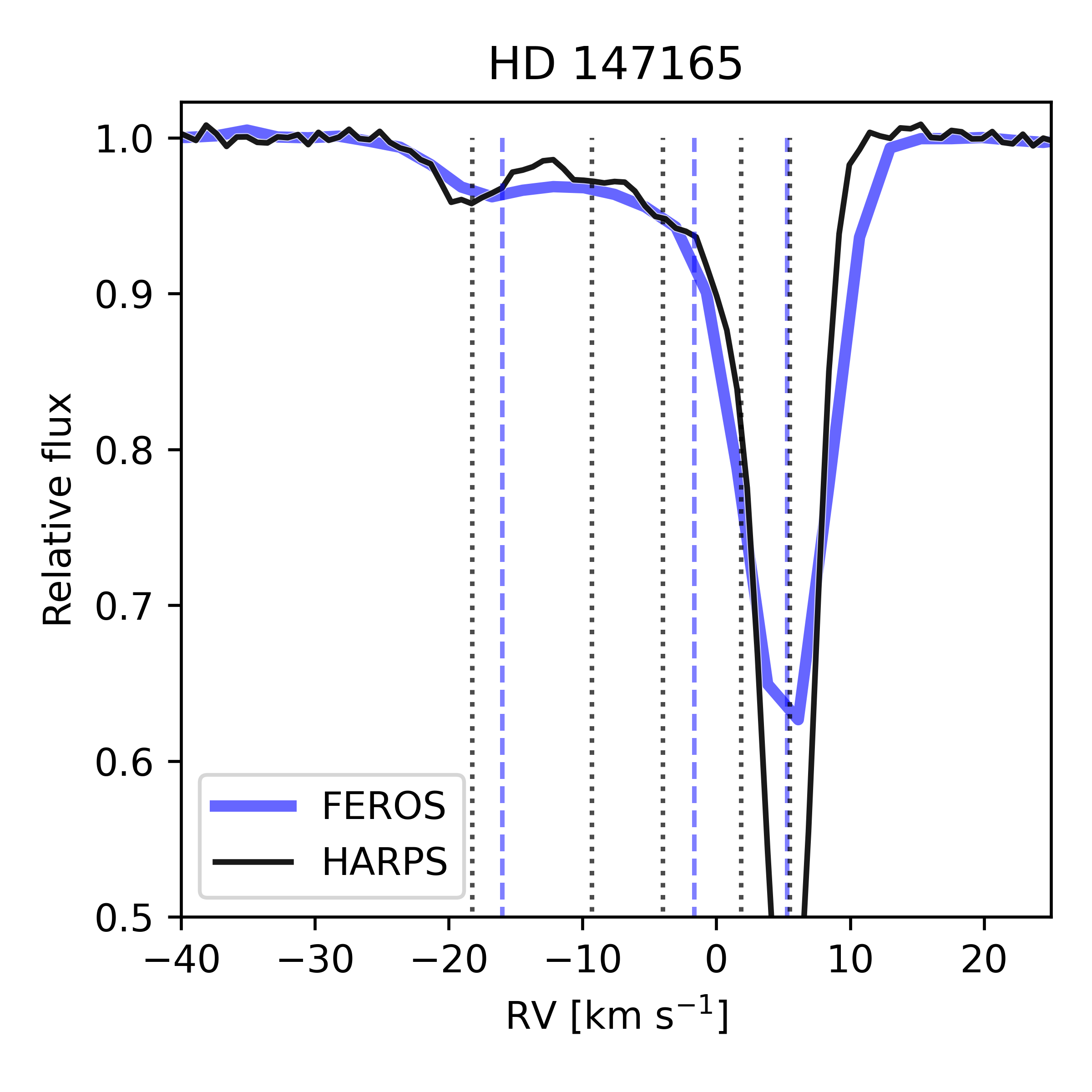}
 \caption{Examples of observed FEROS and HARPS spectra in the line of sight towards HD~147165 ($\sigma$~Sco). The RVs are in LSR. The vertical lines display the velocities of the fitted Gaussian sub-components (dashed for FEROS, dotted for HARPS). It is clear that although the main and the blue components of FEROS can be also identified in HARPS, we lose some kinematic information by working with FEROS. We note that $\sigma$~Sco is one of the few stars for which a triple-component fit would produce a significantly better result than a double-component fit.}
 \label{fig:B}
\end{figure}

We noticed a significant difference from the analysed FEROS spectra in all of the 11 available HARPS spectra. The main issue lies in the presence of unresolved kinematic components. Fig.~\ref{fig:B} shows an example of a line of sight where resolving the kinematic sub-structure would provide additional information about the structure of the intervening ISM. Another interesting example is the $\beta$~Sco system (HD~144217/144218), where a split in the blue-shifted component is missed when using FEROS spectra, producing a systematic offset of a few km\,s$^{-1}$.

\end{appendix}

\end{document}